\documentclass[12pt,a4paper]{article}
\usepackage[a4paper, total={6in, 8in}]{geometry}
\usepackage[utf8]{inputenc}
\usepackage{amsmath}
\usepackage{amsfonts}
\usepackage{amssymb}
\usepackage{cite}
\usepackage{graphicx}
\usepackage{caption}
\usepackage{subcaption}

\begin{document}
\author{Riki Dutta, Sagardeep Talukdar, Gautam Kumar Saharia 
 and \\ Sudipta Nandy \textsuperscript{a}\thanks{email: sudiptanandy@cottonuniversity.ac.in}}
\title{Fokas-Lenells equation dark soliton and gauge equivalent spin equation}
\maketitle

\begin{center}
Abstract
\end{center}

We propose the Hirota bilinearization of the Fokas-Lenells derivative nonlinear Schr\"{o}dinger equation with a non-vanishing background. The bilinear method is applied using an auxilary function to obtain the dark one soliton solution, dark two soliton solution and eventually the scheme for obtaining dark N soliton solutions. The use of auxilary function in bilinearization  makes the method simpler than the ones reported earlier. Later, we have introduced a Lax pair for this integrable equation and using a transformation we have shown that this system is gauge  equivalent to a spin system, namely the Landau-Lifshitz equation.

%%%%%%%%%%%%%%%%%%%%%%%%%%%%%%%%%%%%%%%%%%%%%%

\vspace{5mm}
\section{Introduction}

Fokas-Lenells equation (FLE) \cite{fokas1995class, lenells2009exactly} is an integrable equation which describes the propagation of ultra-short pulses in an optical medium. FLE equation is a special case of a more generalized nonlinear Schr\"{o}dinger equation (NLSE) which includes higher order effects. There are only a few known integrable equations of special cases, namely NLSE \cite{hasegawa1973transmission, hasegawa1973transmission2, serkin2000novel}, higher order NLSE (HNLSE) \cite{hirota1976n, sasa1991new}, derivative NLSE (DNLSE) \cite{kaup1978exact} and the fourth equation is the FLE. A common property in all the equations is the appearance of solitons.  Soliton arise as a result of a balance between the nonlinear and dispersive terms of the wave equations.  While the bright soliton appear as a hump on the  zero background of light, the dark soliton appear as a dip on a continuous background.

The significance of FLE is due to the presence of a spatio-temporal dispersion term along with a cubic nonlinear self-steepening term which account the higher order nonlinear effects in an ultra-short optical pulse. While the study on NLSE, HNLSE and DNLSE is done extensively, the study on FLE is relatively sparse. A few of the recent notable contributions on FLE are solitary wave and elliptical solutions of FLE in presence of perturbation and modulation instability \cite{arshad2019optical}, combined optical solitary waves of FLE \cite{triki2017combined}, dynamical behaviour of soliton solutions of FLE \cite{hendi2021dynamical}, inverse transform of FLE with non zero boundary condition \cite{zhao2021}, derivation of solitons of dimensionless FLE with perturbation term \cite{cinar2022derivation} and more.

%The expressions for soliton have been worked out for the FLE using various methods \cite{matsuno2012direct, matsuno2012direct1, biswas2018optical, jawad2018optical, krishnan2019optical, biswas2018optical1, aljohani2018optical}. These papers have presented the solutions as both dark soliton (dip in a continuous energy background) and anti-dark/bright soliton (bulb in a continuous energy background) depending upon the choice of parameters.
%However, the choice of parameters is limited.
%We here propose the Hirota bilinearisation scheme to obtain the one and multi dark (also anti-dark) soliton solutions of FLE. The proposed scheme includes an auxiliary function making the process simpler and holds less constraints for the parameters.

\vspace{2mm}
The dimensionless expression for FLE \cite{fokas1995class, lenells2009exactly} is
\begin{align} 
\label{FWDU}
iU_t  + a_1 U_{xx} - a_2 U_{xt} + b|U|^2 U + i b a_2 |U|^2U_x = 0
\end{align}

where $U$ is the field function that can describe the waveform of an ultra-short pulse. The suffix $x$ and $t$ denote the partial differentiations with respect to $x$ and $t$ respectively. $U_t$ is the temporal evolution of the pulse, $U_{xx}$ is the group velocity dispersion, $U_{xt}$ is the spatio-temporal dispersion, $|U|^2 U$ is the Kerr nonlinearity effect and $|U|^2 U_x$ is the cubic nonlinear self-steepening effect of the medium.

\vspace{2mm}
Various methods, namely inverse scattering method, Hirota bilinearization and others \cite{lenells2008novel, matsuno2012direct, matsuno2012direct1, biswas2018optical, jawad2018optical, krishnan2019optical, biswas2018optical1, aljohani2018optical, talukdar2023bilinearization, onder2022obtaining, el2023novel, gaballah2023generalized, cinar2022derivation, ullah2023optical} are proposed to obtain the different form of soliton solutions.   
In this manuscript our objective is to propose an alternate simplified and systematic scheme, namely bilinearization by introducing an auxiliary function and obtain a generalized expression for dark and bright dark soliton solutions.

\vspace{2mm}
It is interesting to note that the family of NLSE show gauge equivalence with spin system called Landau-Lifshitz equation (LLE) \cite{takhtajan1979equivalence, kundu1984landau, ghosh1999soliton, ghosh1999inverse}. FLE also belongs to this class of nonlinear Schr\"{o}dinger type equations so it is worth to investigate the gauge connection of FLE with the spin system. To the best of our knowledge in the literature no such work has been reported earlier with FLE.
%The gauge connected forms share the same properties like soliton solutions, conserved quantities, etc. So studying a simpler form like spin form is useful to yield system properties.

\vspace{2mm}
The structure of this manuscript is that in the following section we consider a gauge transformation that will convert eqn. \ref{FWDU} into a simplified form. Then using Hirota bilinearization on this simplified form, we shall derive a dark soliton solution and multi dark soliton solutions and discuss their properties and also mention the condition under which we can get bright soliton solutions. Thereafter in the third section, we propose a gauge transformation of the Lax pair for this FLE to obtain a spin system which is known as the Landau-Lifshitz spin system. Fourth section is the concluding one.

%This equation helps to study the localised waveforms in a nonlinear medium. Optical soliton is a prime example of a localized waveform that the FLE yields \cite{ling2018general, shehata2019new, gomez2022soliton}. The main attraction for the optical solitons is to use it as information carriers in optical fiber based telecommunication for its property to travel a long distance without losing shape. The existence of soliton is a result of balance between dispersion effect and nonlinear effect of the medium. 
%Nonlinear equations like nonlinear Schr\"{o}dinger equation (NLSE), higher order NLSE (HNLSE) also yield soliton solutions \cite{zakharov1974complete, hirota1973exact, sasa1991new}.

\noindent
\section{Bilinearization of FLE with vanishing background} 
Assuming $n=\frac{1}{a_2} $ and $m=\frac{a_1}{a_2} > 0$, consider the gauge transformation 
\begin{align}
\label{GT}
U = \sqrt{\frac{m}{|b|}}n e^{i(n x +2 mn t)} u 
\end{align}
followed by the transformation of variables ($\xi, \tau$)
\begin{align}
\label{VT}
\xi = 2(x+m t), \quad \tau=-\frac{mn^2}{2} t 
\end{align}
we get the following equation
\begin{align} 
\label{FLEsigma}
u_{\xi \tau} = u - 2i \sigma |u|^2 u_\xi      
\end{align}
here $u$ is the field function corresponding to the new transformed system and $\sigma$ = $\frac{b}{|b|}$. Assuming $b$ is positive, we can write eqn. \ref{FLEsigma} as
\begin{align} 
\label{FLE}
u_{\xi \tau} = u - 2i |u|^2 u_\xi      
\end{align}
notice that eqn. \ref{FLE} is the first negative hierarchy of DNLSE. For dark soliton solution of eqn. \ref{FLE} we assume a non vanishing background condition $u \rightarrow \rho\ e^{i \ (\kappa \xi + \omega \tau)}$ as $ \xi \rightarrow \pm \infty $. Under this condition we expect a dark soliton solution with the bilinearization.

\vspace{2mm}
To write  eqn. \ref{FLE} in the bilinear form let us assume 
\begin{align}
\label{bilin0}
u =\frac{g}{f}
\end{align}
where $g$ and  $f$ are two complex functions of ($\xi, \tau$). Consequently  eqn. \ref{FLE} becomes

\begin{multline}
\frac{1}{f^2}(D_\xi D_\tau -1) g.f - \frac{g}{f^3}D_\xi D_\tau(f.f) +  \frac{2i|g|^2}{f^3 f^*}D_\xi (g.f) + \frac{g \lambda f.f}{f^3} - \frac{\lambda g.f}{f^2} +\\
\label{Bilin}
 \frac{s |g|^2}{f^3} - \frac{s |g|^2 f^*}{f^3 f^*} =0
\end{multline} 
where $D_\xi$, $D_\tau $ are  Hirota derivatives \cite{hirota1976n} and are defined as
\begin{align}
D_\xi^m D_\tau^n g(\xi,\tau).f(\xi,\tau)= 
(\frac{\partial}{\partial \xi}  - \frac{\partial}{\partial \xi^\prime})^m
(\frac{\partial}{\partial \tau}  - \frac{\partial}{\partial \tau^\prime})^n
g(\xi,\tau).f(\xi^\prime,\tau^\prime)\Bigg|_{ (\xi=\xi^\prime)(\tau= \tau^\prime)}
\end{align} 

Notice that the last two terms in eqn. \ref{Bilin} contains  an  auxiliary function $s$  which is introduced so that the multilinear eqn. \ref{Bilin} can be cast into two bilinear equations, namely eqns. \ref{BR2} and \ref{BR3}. Here $\lambda$ is a constant to be determined by solving the bilinear eqns. \ref{BR1} - \ref{BR3}. Consequently, the proposed  bilinear equations in terms of $g$, $f$ and $s$  are 
\begin{align}
\label{BR1}
(D_{\xi} D_{\tau} -1 - \lambda)g.f&=0\\
\label{BR2}
(D_{\xi} D_{\tau}  - \lambda)f.f  &= sg^*\\
\label{BR3}
2i D_{\xi} (g.f)   &= sf^* 
\end{align}

To obtain the soliton solution,  $g$ and $f$  are expanded with respect to an arbitrary
parameter $\epsilon$ as follows
\begin{align}
\label{GF}
g&= g_0 \ (1 + \epsilon^2 g_2 + \epsilon^4 g_4 + ... ), \quad \quad \quad 
f= 1 + \epsilon^2 f_2 + \epsilon^4 f_4 + ...
\end{align}
and the auxiliary function $s$ is expanded as  
\begin{align}
\label{S}
s= s_0 \ (1 + \epsilon^2 s_2 + \epsilon^4 s_4 + ... )
\end{align}  

\subsection{Dark one soliton Solution}
The dark one soliton solution (1SS) is obtained by dropping terms of order greater than or equal to  $\epsilon^3 $ in  $g$, $f$ and $s$. Thus from eqn. \ref{bilin0} the dark 1SS of eqn. \ref{FLE} is
\begin{align}
\label{sol1}
u &= \frac{g_0 (1 + \epsilon^2 g_2^{(1)})}{1 + \epsilon^2 f_2^{(1)}}\Big|_{\epsilon=1} 
\end{align}

Let us consider the expressions for $g_0$, $s_0$, $g_2^{(1)}$, $s_2^{(1)}$ and $f_2^{(1)}$ are as follows
\begin{align}
\label{g0}
g_0 &= \rho \ e^{i \ (\kappa \xi + \omega \tau)}\\
\label{s0}
s_0 &= \rho_s \ e^{i \ (\kappa \xi + \omega \tau)} \\
\label{g2}
g_2^{(1)} &= K \ e^{\theta + \theta^*}\\
\label{s2}
s_2^{(1)} &= M \ e^{\theta + \theta^*} \\
\label{f2}
f_2^{(1)} &= T \ e^{\theta + \theta^*}
\end{align}
where $\theta$ = $p \ \xi + \Omega \ \tau$. $p$, $\Omega$, $K$, $M$, $T$ are complex parameters, $\xi$, $\omega$ are real constants and $\rho$, $\rho_s$ are positive constants. Let us consider $p$ = $p_r + i\ p_i$ and $T$ = $T_r + i \ T_i$ where $p_r$, $p_i$, $T_r$ and $T_i$ are real. On substituting the above equations into eqn. \ref{sol1}, we have
\begin{align}
\label{ugf}
u &= \rho \ e^{i \ (\kappa \xi + \omega \tau)} \ \frac{1 + K \  e^{\theta + \theta^*}}{1 + T \ e^{\theta + \theta^*}}
\end{align}
further putting eqns. \ref{g0} - \ref{f2} into eqns. \ref{BR1} - \ref{BR3} yield the following expressions
\begin{align}
\label{rhos}
\rho_s &= - 2\ \kappa \rho\\
\label{lambda}
\lambda &= 2\ \kappa \rho^2\\
\label{omega}
\omega &= -\frac{1}{\kappa} - 2\ \rho^2\\
\label{Omega}
\Omega &= \frac{h}{p}\\
\label{m}
M &= \frac{T^2}{K^*}\\
\label{k}
K &= \gamma \ T^*
\end{align}
where $h$ is real and $\gamma$ is complex and are represented as
\begin{align}
\label{gamma}
\gamma &= \frac{\kappa \ T_i + p_r \ T}{\kappa \ T_i + 
 p_r \ T^*}\\
\label{h}
h &= \frac{|p|^2 \ \kappa \rho^2\ T_i^2}{|p_r\ T + \kappa\ T_i|^2}
\end{align}
and the system obeys the constraint
\begin{align}
\label{cons}
\kappa^2 \ (1 + \kappa \rho^2) T_i^2 + 2 p_r \ \kappa \ (1 + \kappa \rho^2)\ T_i \ T_r + p_r^2 \ |T|^2 &= 0
\end{align}
now keeping one of $T_r$ or $T_i$ fixed the other can be calculated from eqn. \ref{cons}. If we fix $T_r$ then $T_i$ can be expressed as
\begin{align}
\label{Ti}
T_i &= -\frac{\kappa\ (1 + \kappa \rho^2) + \sqrt{-p_r^2 + \kappa^3 \rho^2\ (1 + \kappa \rho^2)}}{p_r^2 + \kappa^2 + \kappa^3 \rho^2}\ p_r\ T_r
\end{align}
from eqn. \ref{Ti} it follows that the condition $\ |p_r|\ \le\ \sqrt{ \kappa^3 \rho^2\ (1 + \kappa \rho^2)}$ must satisfy to obtain 1SS.

\subsubsection{Properties of dark one soliton Solution}

\begin{figure}
	\centering
	\begin{subfigure}{.45\textwidth}
		\includegraphics[width=\textwidth]{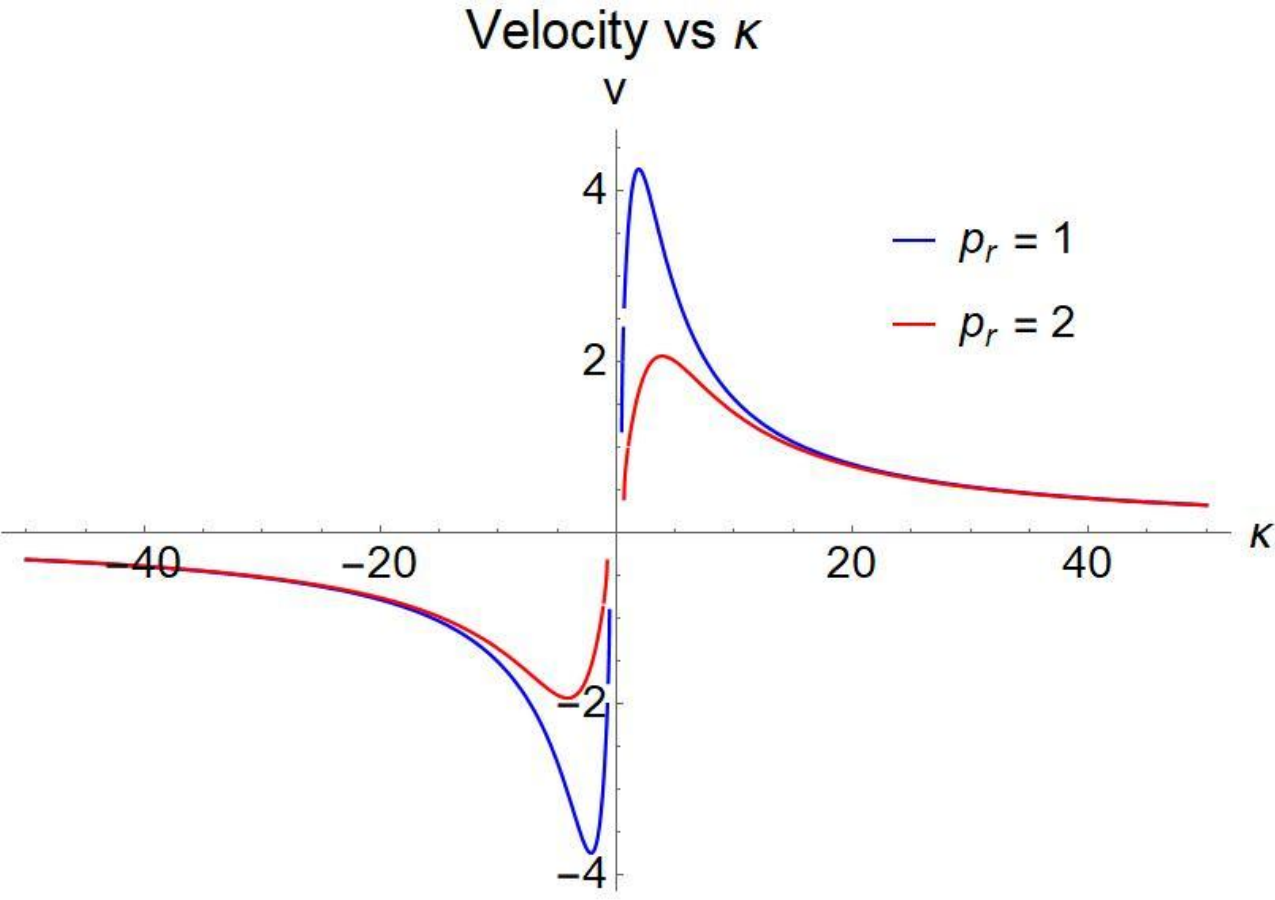}
		\caption{$v$ vs $\kappa$}
		\label{fig11}
	\end{subfigure}
%%%%%%%%%%%%%%
	\begin{subfigure}{.45\textwidth}
		\includegraphics[width=\textwidth]{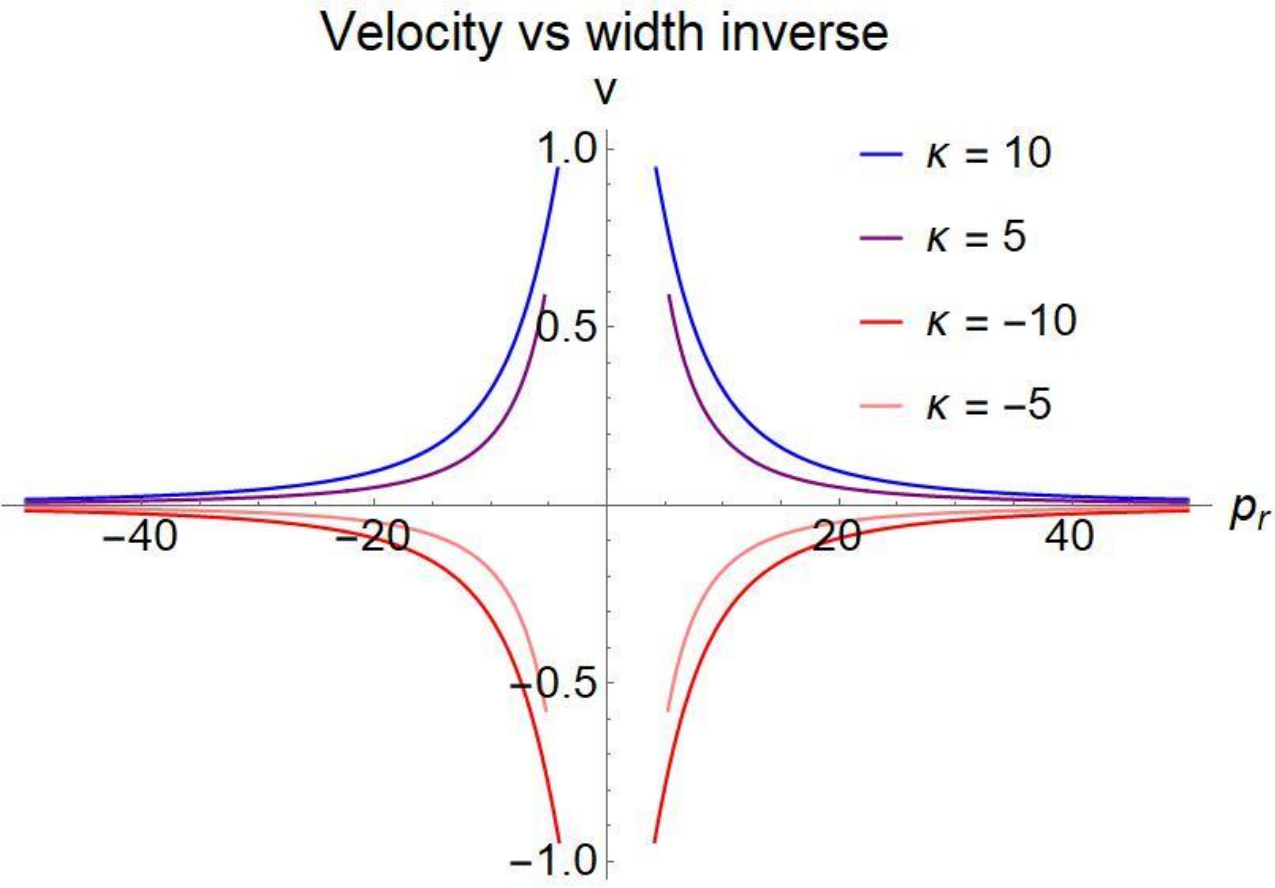}
		\caption{$v$ vs $p_r$}
		\label{fig12}
	\end{subfigure}
	\caption{Variation of velocity (a) with respect to $\kappa$ and fixed $p_r$ = $1$ and $2$ and (b) with respect to $p_r$ and fixed $\kappa$ = $\pm 5$, $\pm 10$. In both the graphs we have fixed $\rho$ = $2$ and $T_r$ can have any value say $\pm 1$, $\pm 2$, etc. the graphs will remain the same.}
\end{figure}

\begin{figure}
	\centering
	\begin{subfigure}{.45\textwidth}
		\includegraphics[width=\textwidth]{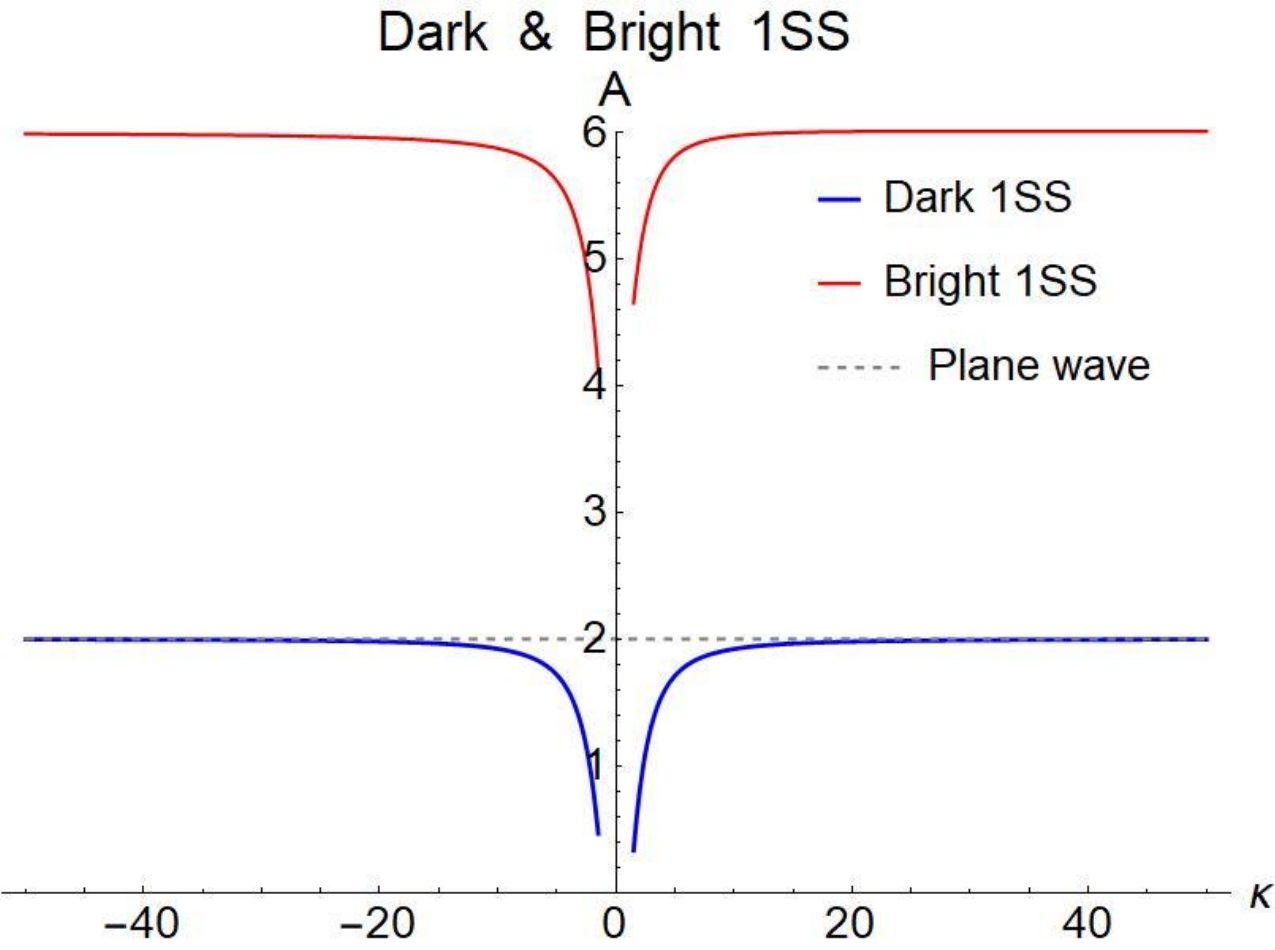}
		\caption{$A$ vs $\kappa$}
		\label{fig21}
	\end{subfigure}
%%%%%%%%%%%%%%
	\begin{subfigure}{.45\textwidth}
		\includegraphics[width=\textwidth]{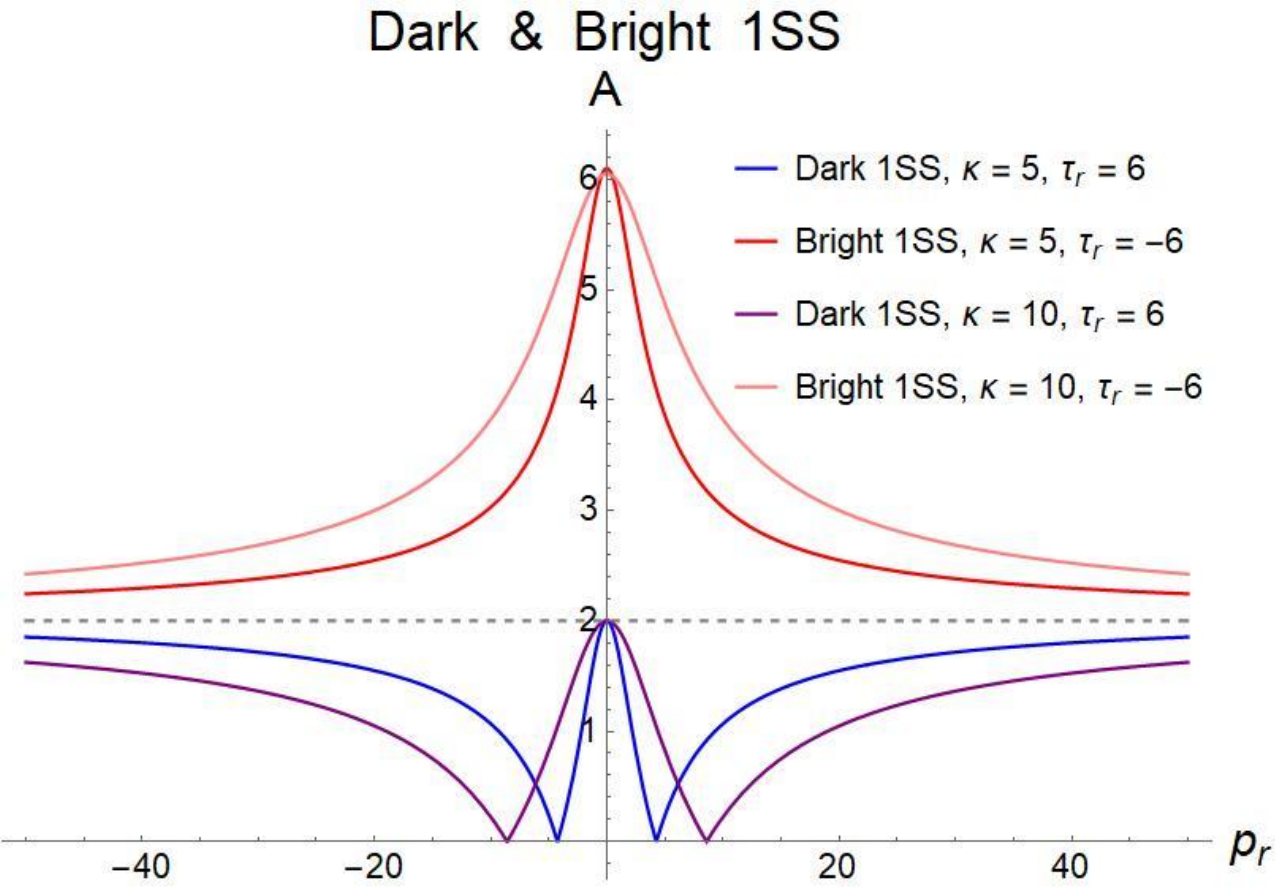}
		\caption{$A$ vs $p_r$}
		\label{fig22}
	\end{subfigure}
	\caption{Variation of amplitude (a) with respect to $\kappa$ and fixed $p_r$ = $1$ and (b) with respect to $p_r$ and fixed $\kappa$ = $\pm 5$, $\pm 10$. In both the graphs we fixed $\rho$ = $2$ and $T_r$ = $6$ and $-6$ for dark and bright soliton respectively.}
\end{figure}

\begin{figure}
	\centering
	\begin{subfigure}{.45\textwidth}
		\includegraphics[width=\textwidth]{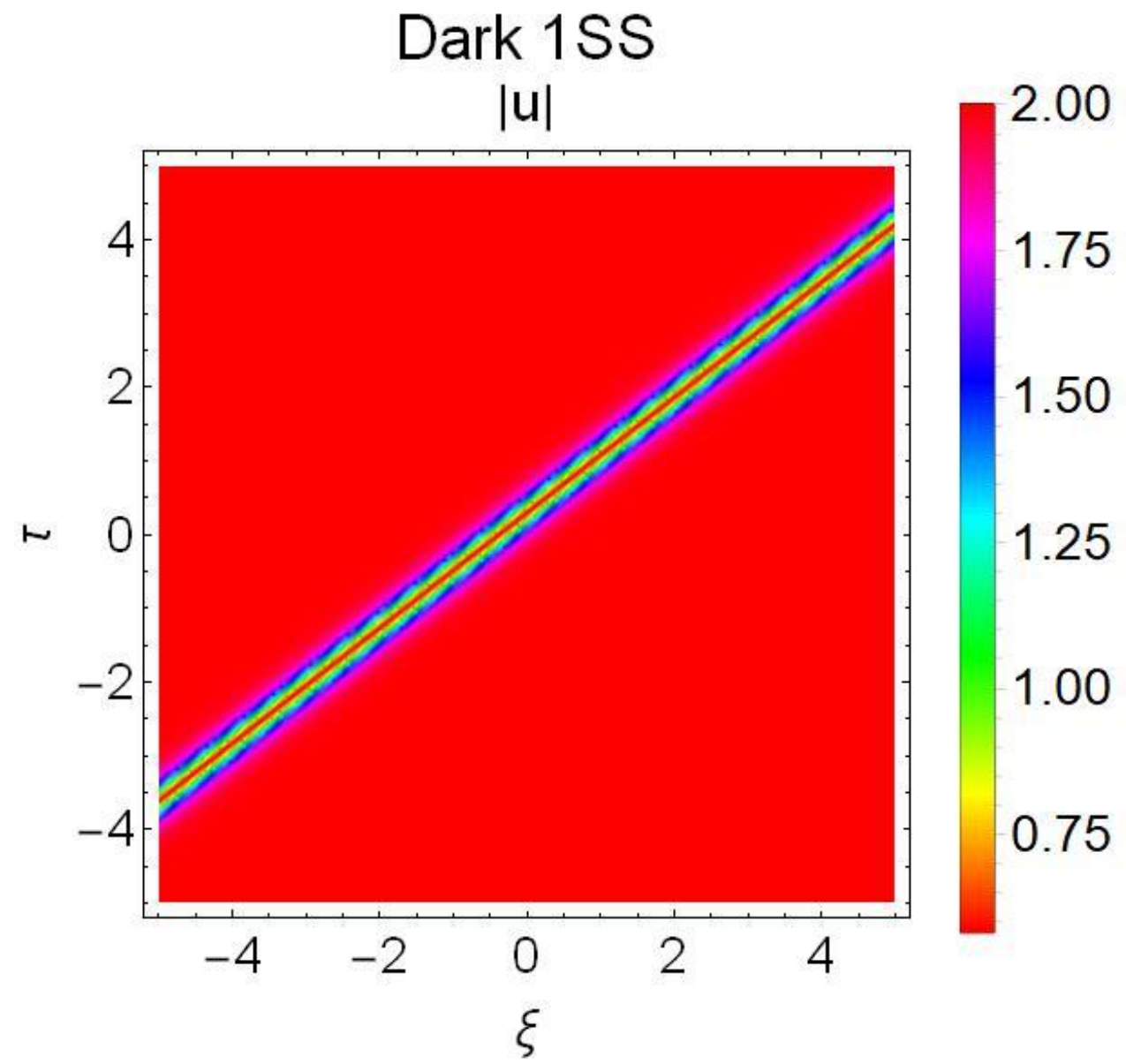}
		\caption{Dark 1SS}
		\label{fig31}
	\end{subfigure}
%%%%%%%%%%%%%%
	\begin{subfigure}{.45\textwidth}
		\includegraphics[width=\textwidth]{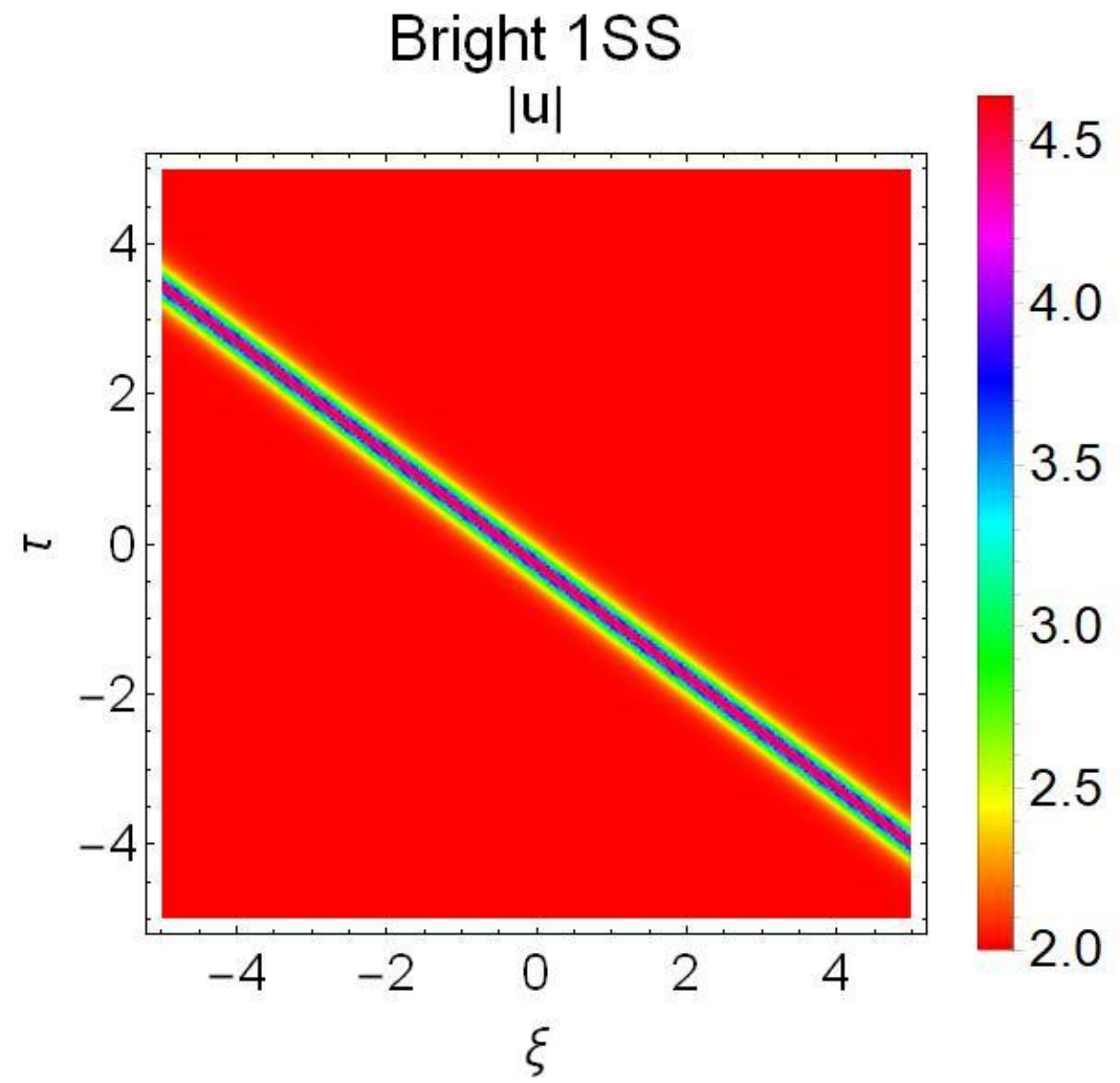}
		\caption{Bright 1SS}
		\label{fig32}
	\end{subfigure}
	\caption{(a) represents dark soliton with $\kappa$ = $-5$ and $T_r$ = $6$ and (b) represents bright soliton with $\kappa$ = $5$ and $T_r$ = $-6$ in density plot. In both the figures we have fixed $\rho$ = $2$.}
\end{figure}

In this part we shall discuss about the properties of 1SS (eqn. \ref{ugf}) namely velocity, width inverse, the criteria upon which the nature of the one soliton obtained will be dark or bright and also discuss the amplitude. One thing to point out that there is always a background present so instead of calling 'bright soliton', an 'anti-dark soliton' term is much more accurate but we shall refer to it as 'bright soliton' for convenience. At first, let us parametrise $\theta$ as
\begin{align}
\theta &= p\ x + \Omega\ t\\
\Rightarrow \theta &= p_r\ (x + v\ t) + i\ p_i\ (x - v\ t)
\end{align}
here $2 p_r$ represents the width inverse and $v$ denotes the velocity of 1SS and is represented as 
\begin{align}
\label{vh}
v &= \frac{h}{p_r^2 + p_i^2}\\
\label{v}
\Rightarrow v &=  \frac{\kappa \rho^2\ T_i^2}{|p_r\ T + \kappa\ T_i|^2}
\end{align}
eqn. \ref{v} shows that $\kappa$ can control both the sign and magnitude of $v$, graphically represented in figure \ref{fig11}. In brief the figure \ref{fig11} interprets that magnitude of $v$ at first increases with the magnitude of $\kappa$ and then decreases. And figure \ref{fig12} shows that the magnitude of $v$ decreases with $p_r$ and the sign of $p_r$ has no effect on $v$, in other words $v$ is symmetric to $p_r$. $T_r$ has no effect on $v$ as $T_i$ is proportional to $T_r$ (from eqn. \ref{Ti}) so $T_r$ is a common term in both numerator and denominator of eqn. \ref{v}. In both the graphs $v$ tends to zero at extreme points of $\kappa$ and $p_r$. These characteristics of $v$ is irrespective to the nature of soliton being dark or bright.
\vspace{2mm}
The amplitude ($A$) of 1SS is expressed as
\begin{align}
\label{A}
A &= \rho\ \Big|\frac{T + \gamma\  |T|}{T + |T|}\Big|
\end{align}
for positive $p_r$, when $\kappa$ and $T_i$ have different signs we get $A$ smaller than $\rho$ and $A$ is bigger than $\rho$ for the same sign of $\kappa$ and $T_i$. For the case of negative $p_r$ the criteria become vice versa. These are the criteria for the soliton to be dark and bright respectively. Figures \ref{fig21} and \ref{fig22} show the variation of $A$ with respect to $\kappa$ and $p_r$ respectively for dark and bright solitons. Figure \ref{fig22} shows that the amplitude of the dark soliton at first decreases then increases with $p_r$ and for bright soliton the amplitude decreases monotonically with $p_r$ and in both dark and bright cases the amplitude $A \rightarrow \rho$ as $p_r \to \pm \sqrt{ \kappa^3 \rho^2\ (1 + \kappa \rho^2)}$. One important fact to notice is that in the limit $p_r \to 0$, the dark soliton reduces to a plane wave that means algebraic dark soliton does not exist. However algebraic bright soliton exists. Figure \ref{fig22} shows the same. In the same graph we see that at certain values of $p_r$ the amplitude of dark soliton reduces to zero indicating the occurrence of a black soliton. Figures \ref{fig31} and \ref{fig32} represent the propagation of dark and bright soliton respectively in 2D density plot.

%%%%%%%%%%%%%%%%%%%%

\subsection{Dark two soliton Solution}
The dark two soliton solution (2SS) is obtained by dropping the terms of order greater than or equal to  $\epsilon^5 $ in  $g$, $f$ and $s$. Thus from eqn. \ref{bilin0} the dark 2SS of eqn. \ref{FLE} is

\begin{align}
\label{sol2}
u &= \frac{g_0 (1 + \epsilon^2 g_2^{(2)} + \epsilon^4 g_4^{(2)})}{1 + \epsilon^2 f_2^{(2)} + \epsilon^4 f_4^{(2)}}\Big|_{\epsilon=1} 
\end{align} 
Let us consider the expressions for $g_2^{(2)}$, $g_4^{(2)}$,$s_2^{(2)}$, $s_4^{(2)}$, $f_2^{(2)}$ and $f_4^{(2)}$ in 2SS are as follows
\begin{align}
\label{g22}
g_2^{(2)} &= K_1\ e^{\theta_1 + \theta_1^*} + K_2\ e^{\theta_2 + \theta_2^*} \\
\label{g4}
g_4^{(2)} &= K_{12}\ e^{\theta_1 + \theta_1^* + \theta_2 + \theta_2^*} \\
\label{s22}
s_2^{(2)} &= M_1\ e^{\theta_1 + \theta_1^*} + M_2\ e^{\theta_2 + \theta_2^*} \\
\label{s4}
s_4^{(2)} &= M_{12}\ e^{\theta_1 + \theta_1^* + \theta_2 + \theta_2^*} \\
\label{f22}
f_2^{(2)} &= T_1\ e^{\theta_1 + \theta_1^*} + T_2\ e^{\theta_2 + \theta_2^*}\\
\label{f4}
f_4^{(2)} &= T_{12}\ e^{\theta_1 + \theta_1^* + \theta_2 + \theta_2^*}
\end{align}
where $\theta_j$ = $p_j\ \xi + \Omega_j\ \tau$. $p_j$, $\Omega_j$, $K_j$, $M_j$, $T_j$, $K_{12}$, $M_{12}$, $T_{12}$ are complex parameters ($j = 1,2$). Let us consider $p_j$ = $p_{r_j} + i\ p_{i_j}$ and $T_j$ = $T_{r_j} + i \ T_{i_j}$ where $p_{r_j}$, $p_{i_j}$, $T_{r_j}$ and $T_{i_j}$ are real. On substituting the above equations into eqn. \ref{sol2}, we have
\begin{align}
\label{ugf2}
u &= \rho\ e^{i\ (\kappa \xi + \omega \tau)} \frac{1 + K_1\ e^{\theta_1 + \theta_1^*} + K_2\ e^{\theta_2 + \theta_2^*} + K_{12}\ e^{\theta_1 + \theta_1^* + \theta_2 + \theta_2^*}}{1 + T_1\ e^{\theta_1 + \theta_1^*} + T_2\ e^{\theta_2 + \theta_2^*} + T_{12}\ e^{\theta_1 + \theta_1^* + \theta_2 + \theta_2^*}}
\end{align}
now putting eqns. \ref{g22} - \ref{f4} into eqns \ref{BR1} - \ref{BR3} yield the following expressions:
\begin{align}
\label{Omega2}
\Omega_j &= \frac{h_j}{p_j}\\
\label{m2}
M_j &= \frac{T_j^2}{K_j^*}\\
\label{m12}
M_{12} &= \frac{T_{12}^2}{K_{12}^*}\\
\label{k2}
K_j &= \gamma_j \ T_j^*\\
\label{k12}
K_{12} &= \gamma_1\ \gamma_2 \ T_{12}^*\\
\label{t12}
T_{12} &= c \ T_1\ T_2
\end{align}
where $h_j$'s are real and $\gamma_j$'s are complex and are represented as
\begin{align}
\label{gamma2}
\gamma_j &= \frac{\kappa \ T_{i_j} + p_{r_j} \ T_j}{\kappa \ T_{i_j} + p_{r_j} \ T_j^*}\\
\label{h2}
h_j &= \frac{|p_j|^2 \ \kappa \rho^2\ T_{i_j}^2}{|p_{r_j}\ T_j + \kappa\ T_{i_j}|^2}
\end{align}
and $c$ is real expressed as
\begin{align}
c &= \frac{(p_{1_r}\ T_{1_r}\ T_{2_i} - 
    p_{2_r}\ T_{1_i}\ T_{2_r})^2 + (p_{1_r} - p_{2_r})^2\ T_{1_i}^2\ T_{2_i}^2}{(p_{1_r}\ T_{1_r}\ T_{2_i} - 
    p_{2_r}\ T_{1_i}\ T_{2_r})^2 + (p_{1_r} + p_{2_r})^2\ T_{1_i}^2\ T_{2_i}^2}
\end{align}
the system obeys the constraints
\begin{align}
\label{cons2}
\kappa^2 \ (1 + \kappa \rho^2) T_{i_j}^2 + 2 p_{r_j} \ \kappa \ (1 + \kappa \rho^2)\ T_{i_j} \ T_{r_j} + p_{r_j}^2 \ |T_j|^2 &= 0
\end{align}
if we fix $T_{r_j}$ then $T_{i_j}$ can be calculated from eqn. \ref{cons2} as
\begin{align}
\label{Ti2}
T_{i_j} &= -\frac{\kappa\ (1 + \kappa \rho^2) + \sqrt{-p_{r_j}^2 + \kappa^3 \rho^2\ (1 + \kappa \rho^2)}}{p_{r_j}^2 + \kappa^2 + \kappa^3 \rho^2}\ p_{r_j}\ T_{r_j}
\end{align}
from eqn. \ref{Ti2}, the conditions $\ |p_{r_j}|\ \le\ \sqrt{ \kappa^3 \rho^2\ (1 + \kappa \rho^2)}$ must satisfy to obtain 2SS.

\subsubsection{Properties of dark two soliton solution}

In a similar way as we have done in section 2.1.1, to investigate the properties of 2SS we first parametrise $\theta_j$ ($j=1,2$) as
\begin{align}
\theta_j &= p_j\ x + \Omega_j\ t\\
\Rightarrow \theta_j &= p_{r_j}\ (x + v_j\ t) + i\ p_{i_j}\ (x - v_j\ t)
\end{align}
where $v_j$ denotes the velocity of the jth soliton and represented as 
\begin{align}
\label{vh2}
v_j &= \frac{h_j}{p_{r_j}^2 + p_{i_j}^2}\\
\label{v2}
\Rightarrow v_j &=  \frac{\kappa \rho^2\ T_{i_j}^2}{|p_{r_j}\ T_j + \kappa\ T_{i_j}|^2}
\end{align}
and $2 p_{r_j}$ represents the width inverse of the jth soliton of 2SS. The amplitude, $A_j$ of the jth soliton is given as
\begin{align}
\label{A2}
A_j &= \rho\ \Big|\frac{T_j + \gamma_j\  |T_j|}{T_j + |T_j|}\Big|
\end{align}

\begin{figure}
	\centering
	\begin{subfigure}{.45\textwidth}
 		\includegraphics[width=\textwidth]{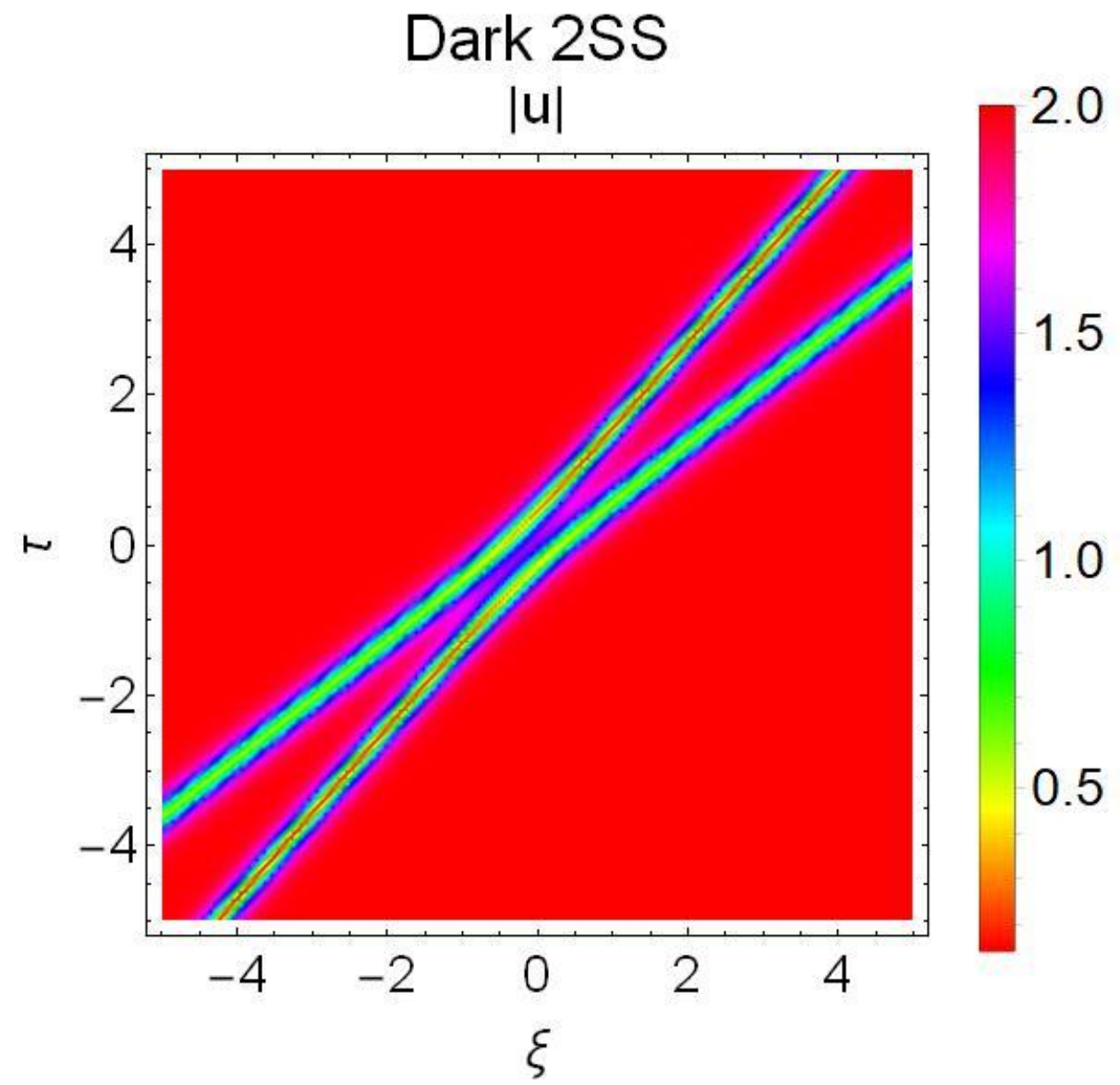}
		\caption{Dark-dark 2SS}
		\label{fig41}
	\end{subfigure}
%%%%%%%%%%%%%%
	\begin{subfigure}{.45\textwidth}
		\includegraphics[width=\textwidth]{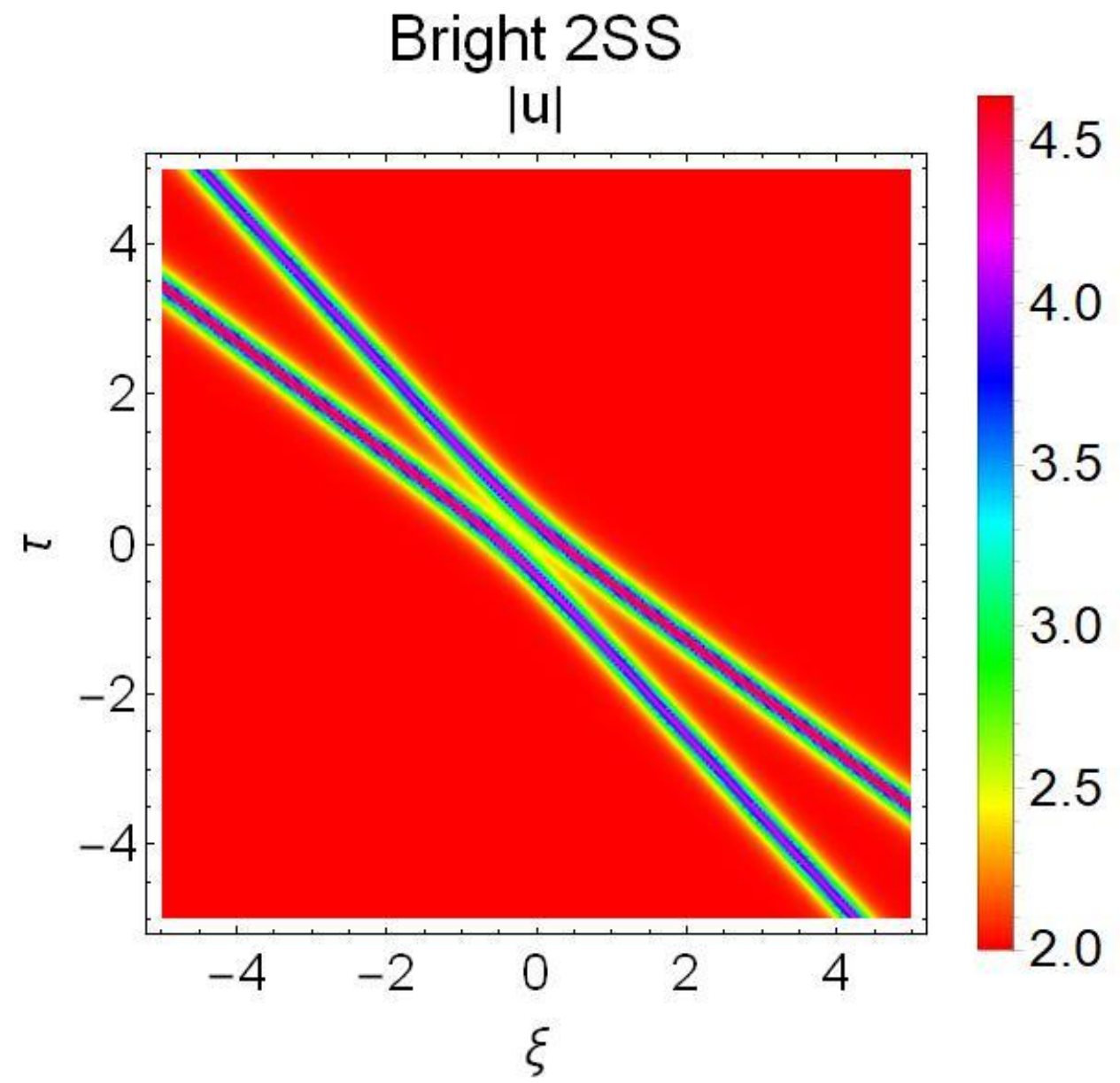}
		\caption{Bright-bright 2SS}
		\label{fig42}
	\end{subfigure}
	\caption{(a) represent the dark-dark 2SS, here $p_{r_1}$ = $3$, $p_{r_2}$ = $4$, $T_{r_1}$ = $6$, $T_{r_2}$ = $9$ and $\kappa$ = $-5$ (b) represent the bright-bright 2SS, here $p_{r_1}$ = $3$, $p_{r_2}$ = $4$, $T_{r_1}$ = $-6$, $T_{r_2}$ = $-9$ and $\kappa$ = $5$. In both the graphs we fixed $\rho = 2$.}
\end{figure}

The velocity and amplitude of the jth soliton obey the same characteristics as that of 1SS obey as discussed in the section 2.1.1. Interaction of two individual solitons of 2SS is demonstrated in figures \ref{fig41} and \ref{fig42} for a dark-dark case and bright-bright case respectively in 2D density plot.

\subsubsection{Asymptotic analysis of 2SS}

To study the amplitudes and the phase shift of 2SS we shall perform asymptotic analysis. When asymptotically apart from each other, 2SS is essentially two separated 1SS. In the limit of before interaction $\tau \to -\infty$, if we fix $\theta_1$, this implies $|e^{\theta_1}|$ is finite and $|e^{\theta_2}| \to \infty$. This corresponds to an asymptotic form of eqn. \ref{ugf2} which is expressed as
\begin{align}
\label{ua1kt}
u_{1-} &=  \rho\ e^{i\ (\kappa \xi + \omega \tau)}\ \frac{K_2\ (1 + c\ K_1\ e^{\theta_1 + \theta_1^*})}{T_2\ (1 + c\ T_1\ e^{\theta_1 + \theta_1^*})}\\
\label{ua1}
\Rightarrow u_{1-} &=  \rho\ e^{i\ (\kappa \xi + \omega \tau)}\ e^{i\ \phi}\ \frac{(1 + K_1^\prime\ e^{\theta_1 + \theta_1^*})}{(1 + T_1^\prime\ e^{\theta_1 + \theta_1^*})}
\end{align}
here $ K_1^\prime = c\ K_1$ and  $T_1^\prime = c\ T_1$. Eqn. \ref{ua1} is nothing but 1SS (from eqn. \ref{ugf}) with an additional phase $\phi$. The arise of $\phi$ is due to term $\frac{K_2}{T_2}$ in eqn. \ref{ua1kt} and the expression of $\phi$ is given in a few steps below (in eqn. \ref{phi12}). \\
In the limit of after interaction $\tau \to \infty$, again we have fixed $\theta_1$ then this will imply $|e^{\theta_1}|$ is finite and $|e^{\theta_2}| \to 0$. The corresponding asymptotic form of eqn. \ref{ugf2} expressed as
\begin{align}
\label{ua2}
u_{1+} &=  \rho\ e^{i\ (\kappa \xi + \omega \tau)}\ \frac{(1 + K_1\ e^{\theta_1 + \theta_1^*})}{(1 + T_1\ e^{\theta_1 + \theta_1^*})}
\end{align}
we see eqn. \ref{ua2} is same as eqn. \ref{ugf} with no additional phase. The amplitude of both the eqns. \ref{ua1} and \ref{ua2} is same and can be expressed as
\begin{align}
\label{amp12}
A_{asymp} &= \rho\ \sqrt{\frac{(T_1+T_1^*)(K_1+K_1^*)-2(|T_1|^2+|K_1|^2) + 2\ \sigma_1 |K_1 - T_1||K_1 - T_1^*|}{(T_1-T_1^*)^2}}
\end{align}
where $\sigma_1$ = sign($T_{r_1}$). This is one of the important characteristics of a soliton that upon interaction the amplitude of the soliton remains same only a phase shift occurs. In this case the phase shift is $\phi$ and its value is given by
\begin{align}
\label{phi12}
\phi &= tan^{-1}\Big[-\frac{2 \kappa\ T_{i_2}^2\ (\kappa\ T_{r_2}\ T_{i_2} + p_{r_2} |T_2|^2)}{(\kappa\ (T_{r_2} - T_{i_2})\ T_{i_2} + p_{r_2} |T_2|^2)\ (\kappa\ (T_{r_2} + T_{i_2})\ T_{i_2} + p_{r_2} |T_2|^2)}\Big]
\end{align}

We can perform the same analysis by keeping $\theta_2$ fixed. In the limit $\tau \to -\infty$, the asymptotic form of eqn. \ref{ugf2} becomes
\begin{align}
\label{ua3}
u_{2-} &=  \rho\ e^{i\ (\kappa \xi + \omega \tau)}\ \frac{(1 + K_2\ e^{\theta_2 + \theta_2^*})}{(1 + T_2\ e^{\theta_2 + \theta_2^*})}
\end{align}
and in the limit $\tau \to \infty$, the asymptotic form of eqn. \ref{ugf2} becomes
\begin{align}
\label{ua4}
u_{2+} &=  \rho\ e^{i\ (\kappa \xi + \omega \tau)}\ \frac{K_1\ (1 + c\ K_2\ e^{\theta_2 + \theta_2^*})}{T_1\ (1 + c\ T_2\ e^{\theta_2 + \theta_2^*})}
\end{align}
The amplitude of both $u_{2\pm}$ can be expressed as
\begin{align}
\label{amp34}
A_{asymp}^{\prime} &= \rho\ \sqrt{\frac{(T_2+T_2^*)(K_2+K_2^*)-2(|T_2|^2+|K_2|^2) + 2\ \sigma_2 |K_2 - T_2||K_2 - T_2^*|}{(T_2-T_2^*)^2}}
\end{align}
where $\sigma_2$ = sign($T_{r_2}$). The phase shift in this case is expressed as
\begin{align}
\phi^{\prime} &= tan^{-1}\Big[-\frac{2 \kappa\ T_{i_1}^2\ (\kappa\ T_{r_1}\ T_{i_1} + p_{r_1} |T_1|^2)}{(\kappa\ (T_{r_1} - T_{i_1})\ T_{i_1} + p_{r_1} |T_1|^2)\ (\kappa\ (T_{r_1} + T_{i_1})\ T_{i_1} + p_{r_1} |T_1|^2)}\Big]
\end{align}

\subsection{Scheme for dark N soliton solution}

To obtain dark N soliton solution (NSS), we have to drop the terms greater than or equal to $\epsilon^{2N+1}$ in $g$, $f$ and $s$. Eqn. \ref{bilin0} becomes
\begin{align}
\label{un}
u &= \frac{g_0 (1 + \epsilon^2 g_2^{(N)} + . . . + \epsilon^{2N} g_{2N}^{(N)})}{1 + \epsilon^2 f_2^{(2)} + . . . + \epsilon^{2N} f_{2N}^{(N)}}\Big|_{\epsilon=1}
\end{align}
here the terms in $g^{(N)}$ and $f^{(N)}$ can be expressed as
\begin{align}
\label{gn}
g^{(N)} &= g_0\ \Big(1 + \sum_{j=1}^N\ K_j\ e^{\theta_j + \theta_j^*}  + \frac{1}{2} \sum_{j,l=1;\ j\neq l}^N K_{jl}\ e^{\theta_j + \theta_j^* + \theta_l + \theta_l^*} + ... + K_{12...N}\ e^{\ \sum\limits_{j=1}^N \theta_j + \theta_j^*} \Big) \\
\label{fn}
f^{(N)} &= 1 + \sum_{j=1}^N\ T_j\ e^{\theta_j + \theta_j^*}  + \frac{1}{2} \sum_{j,l=1;\ j\neq l}^N T_{jl}\ e^{\theta_j + \theta_j^* + \theta_l + \theta_l^*} + ... + T_{12...N}\ e^{\ \sum\limits_{j=1}^N \theta_j + \theta_j^*}
\end{align}
where $\theta_j = p_j\ x + \Omega_j\ t$. $p_j$, $\Omega_j$, $K_j$, $K_{jl}$, ..., $K_{12...N}$, $T_j$, $T_{jl}$, ..., $T_{12...N}$,  are complex parameters (subscripts go from $1$ to $N$). These parameters can be calculated using the bilinear eqns. \ref{BR1} - \ref{BR3}. The N soliton system consists of N number of constraints just like 2SS have two constraints expressed in eqn. \ref{cons2}. The individual jth soliton of NSS have velocity expression $v_j$ from eqn. \ref{v2} and amplitude expression $A_j$ from eqn. \ref{A2} and obey the same characteristics with respect to the parameters $\rho$, $\kappa$, $p_{r_j}$, $T_j$ as discussed in the section 2.1.1.

\section{Gauge Connected Landau-Lifshitz equation} 

In this section we derive the gauge connected LLE of eqn. \ref{FLE} by exploiting the equivalence of the Lax pairs of the two gauge connected systems. Let us consider the Lax pair ($L$, $M$) for eqn. \ref{FLE}
\begin{align}
\label{LaxFLx}
\Psi_{\xi} = L \Psi \\
\label{LaxFLt}
\Psi_{\tau} = M \Psi
\end{align}

where $\Psi$ is the Jost function corresponding to the field function $u$ and ($L$, $M$) are given by 
\begin{align}
\label{LaxPairL}
L = \frac{-i\zeta^2 }{2} \Sigma   + \zeta u_{\xi}          \\
\label{LaxPairM}
M= \frac{i}{2 \zeta^2 } \Sigma - \frac{i}{\zeta}\Sigma u + i\Sigma u^2
\end{align} 

$\zeta$ is the spectral parameter and $\Sigma$ and $u$ are $2\times 2 $ matrices given as follow

\vspace{4mm}
$ \quad \Sigma = \left(  \begin{tabular}{c c}
%\hline 
1 & 0 \\ 
%\hline 
0 & -1  \\ 
%\hline 
\end{tabular} \right), \quad \quad u = \left(  \begin{tabular}{c c}
%\hline 
0 & q \\ 
%\hline 
$-q^*$ & 0   \\
%\hline 
\end{tabular} \right) $ \\

\vspace{1mm}
from eqns. \ref{LaxFLx} and \ref{LaxFLt}, we can write $(\Psi_{\xi})_{\tau} = (\Psi_{\tau})_{\xi}$, this gives:
\begin{align}
\label{ComEqn}
L_{\tau} - M_{\xi} + [L, M] = 0
\end{align}

eqn. \ref{ComEqn} is the compatibility condition also called zero curvature equation. Under a local gauge transformation we write a matrix $g$ such that:\\

$\quad g({\xi}, {\tau}, \zeta_0) = \Psi ({\xi}, {\tau}, \zeta) |_{\zeta=\zeta_0}$ \\

the Jost function $\Psi$ changes to $\Phi$ as:\\

$\quad \Psi \rightarrow \Phi ({\xi}, {\tau}, \zeta, \zeta_0) = g({\xi}, {\tau}, \zeta_0)^{-1} \Psi ({\xi}, {\tau}, \zeta)$\\

$\Phi$ is the Jost function corresponding to the spin field function $S$ of Landau-Lifshitz (LL) system. The new Lax pair ($L^\prime$, $M^\prime$) associated with $\Phi$ expressed as:
\begin{align}
\label{LaxFL2x}
\Phi_{\xi} = L^\prime \Phi \\
\label{LaxFL2t}
\Phi_{\tau} = M^\prime \Phi
\end{align}

and are related to ($L$, $M$) as:
\begin{align}
\label{LaxPairL2}
L^\prime = g^{-1} L g - g^{-1} g_{\xi} \\
\label{LaxPairM2}
M^\prime = g^{-1} M g - g^{-1} g_{\tau}
\end{align}

but
\begin{align}
\label{gxgt}
g_{\xi} g^{-1} = L_0, \quad \quad \quad  g_{\tau} g^{-1} = M_0
\end{align}

therefore,
\begin{align}
\label{LPL2}
L^\prime = g^{-1} (\frac{-i}{2} (\zeta^2 - \zeta_0^2) \Sigma + (\zeta - \zeta_0) u_{\xi}) g \\
\label{LPM2}
M^\prime = g^{-1} (\frac{i}{2} (\frac{1}{\zeta^2} - \frac{1}{\zeta_0^2}) \Sigma - i (\frac{1}{\zeta} - \frac{1}{\zeta_0})\Sigma u) g
\end{align}

where $L_0$ and $M_0$ are:
\begin{align}
\label{L0}
L_0 = L |_{\zeta = \zeta_0} = \frac{-i\zeta_0^2 }{2} \Sigma + \zeta_0 u_{\xi} \\
\label{M0}
M_0 = M |_{\zeta = \zeta_0} = \frac{i}{2 \zeta_0^2 } \Sigma - \frac{i}{\zeta_0}\Sigma u + i\Sigma u^2
\end{align}

the compatibility condition eqn. \ref{ComEqn} in terms of the new Lax pair ($L^\prime$, $M^\prime$) takes the form:
\begin{align}
\label{ComEqn2}
L^\prime_{\tau} - M^\prime_{\xi} + [L^\prime, M^\prime] = 0
\end{align}

this is the compatibility condition corresponding to $S$. $S$ being spin field function of LL system can be represented in terms of $g$ as:
\begin{align}
\label{S}
S = g^{-1} \Sigma g
\end{align}

and satisfies
\begin{align}
\label{S2}
S^2 = I
\end{align}

using eqn. \ref{gxgt}, the derivatives of $S$ namely $S_{\xi}$, $S_{\tau}$ and also the terms $S S_{\xi}$, $S S_{\tau}$ can be written as:
\begin{align}
S_{\xi} = g^{-1} [\Sigma, L_0] g
\label{Sx}
\Rightarrow S_{\xi} = 2 \zeta_0 g^{-1} \Sigma u_{\xi} g\\
S_{\tau} = g^{-1} [\Sigma, M_0] g
\label{St}
\Rightarrow S_{\tau} = - \frac{2 i}{\zeta_0} g^{-1} u g\\
\label{SSx}
S S_{\xi} = 2 \zeta_0 g^{-1} u_{\xi} g \\
\label{SSt}
S S_{\tau} = - \frac{2 i}{\zeta_0} g^{-1} \Sigma u g
\end{align}

again using the above four expressions in eqns. \ref{LPL2} and \ref{LPM2}, we get:
\begin{align}
\label{L2S}
L^\prime = \frac{-i}{2} (\zeta^2 - \zeta_0^2) S + (\zeta - \zeta_0)  \frac{1}{2 \zeta_0} SS_{\xi} \\
\label{M2S}
M^\prime = \frac{i}{2} (\frac{1}{\zeta^2} - \frac{1}{\zeta_0^2}) S + (\frac{1}{\zeta} - \frac{1}{\zeta_0}) \frac{\zeta_0}{2} SS_{\tau}
\end{align}

hence, the compatibility condition eqn. \ref{ComEqn2} yields:
\begin{align}
\label{Gauge}
S_{\xi} + \frac{\zeta_0^2}{2 i} [S, S_{{\tau}{\xi}}] = 0
\end{align}

this is the gauge connected LLE for the eqn. \ref{FLE}. One important thing to note is that eqn. \ref{Gauge} is consistent for the system of eqns. \ref{LaxFL2x}, \ref{LaxFL2t}, \ref{L2S}, \ref{M2S} in any matrix $S$ of arbitrary dimension as long as eqn. \ref{S2} holds.

%%%%%%%%%%%%%%%%%%%%		Conclusion		%%%%%%%%%%%%%%%%%%%%

\section{Conclusion}

We have bilinearized the Fokas-Lenells derivative nonlinear Schr\"{o}dinger equation (FLDE) with a non-vanishing boundary condition. In the proposed bilinearization we have used an auxiliary function to derive explicitly the one dark soliton solution and two dark soliton solution and represent the scheme for obtaining N soliton solution. We have also discussed the criteria for the soliton to become dark or bright. Our result accepts a wide range of values for the parameters ($K$'s, $T$'s) unlike the results published earlier. We have obtained a gauge equivalence between FLDE and Landau-Lifshitz (LL) spin system with the explicit construction of the new equivalent Lax pair. We believe that the derived dark (and bright) soliton solutions will be useful in optical communication and other nonlinear fields of physics where higher order effects like spatio-temporal evolution and cubic nonlinear self-steepening effects are taken into account. The gauge equivalent LL equation will be useful to study the integrability properties of the FLDE.

%%%%%%%%%%%%%%%%%%%%		ACKNOWLEDGEMENT		%%%%%%%%%%%%%%%%%%%%

\section*{Acknowledgement}
R Dutta and S Talukdar acknowledge DST, Govt. of India for INSPIRE fellowship award. Corresponding award number DST/INSPIRE Fellowship/2020/IF200303 and DST/INSPIRE Fellowship/2020/IF200278.

\bibliographystyle{elsarticle-num}
\bibliography{FLEDarkLLE}

\end{document}